\documentclass[12pt]{iopart}
\pdfoutput=1 
\usepackage{graphicx}
\begin{document}

\title[Pulse power supply for the horns of CERN to Fr\'ejus super beam]  {Study of the pulse power supply unit for the four-horn system of the CERN to Fr\'ejus neutrino super beam}

\author{E Baussan, E Bouquerel, M Dracos, G Gaudiot, F Osswald, \\P Poussot, N Vassilopoulos, J Wurtz and V Zeter}

\address{Institut Pluridisciplinaire Hubert Curien (IPHC),\\CNRS, 23 rue du loess, 67037 Strasbourg cedex 2, France}


\ead{ppoussot@iphc.cnrs.fr corresponding author}

\begin{abstract} 
The power supply studies for the four-horn system for the CERN to Fr\'ejus neutrino Super Beam oscillation experiment are discussed here. The power supply is being studied to meet the physics potential and the mega-watt (MW) power requirements of the proton driver of the Super Beam. A one-half sinusoid current waveform with a 350~kA maximum current and pulse length of 100~$\mu$s at 50~Hz frequency is generated and distributed to four-horns. In order to provide the necessary current needed to focus the charged mesons producing the neutrino beam, a bench of capacitors is charged  at 50~Hz frequency to a +12~kV reference voltage and then discharged through a large switch to each horn via a set of strip-lines at the same rate. A current recovery stage allows to invert rapidly the negative voltage of the capacitor after the discharging stage in order to recuperate large part of the injected energy and thus to limit the power consuption. The energy recovery efficiency of that system is very high at 97~\%. For feasibility reasons, a modular architecture has been adopted with 8 modules connected in parallel to deliver 44~kA peak currents into the four-horn system. 
\end{abstract}


\vspace{2pc}

\maketitle

\section{Introduction}

The European Design Study EUROnu \cite{designreport} has investigated the possibility of using the CERN superconductive proton linac (SPL) \cite{spl1, spl2} to produce a neutrino Super Beam. After the high power SPL, an accumulator ring is foreseen to compress the proton-bunches from 500~$\mu$s to 3.5~$\mu$s. The proton-beam is foreseen to be separated by a series of kicker magnets into four beam lines. At the end,  each of the four-beams will be focused by a series of quadruples and correctors to a horn/target assembly to focus mesons along the direction of the detector thus producing the neutrino beam. The four-horn/target assembly is chosen to accommodate the heat produced by the 4~MW proton beam power. The four-horn/target system will be placed within a single large helium vessel. The downstream part of the neutrino beam  consists of several collimators, the steal decay tunnel where the neutrinos are produced by mesons decays and the graphite hadron-beam absorber at the end \cite{designreport, elian}.

Each magnetic horn is crossed by high current to produce the necessary magnetic field needed for the meson focusing. The horn is characterized by very low inductance and resistance values of 0.9 $\mu$H and 0.235 m$\Omega$ respectively. The four-horn system has to be pulsed at 50~Hz by a half-sinusoid waveform of 100~$\mu$s and 350~kA peak current.  In nominal operation mode, each horn works at 12.5~Hz frequency and 9~kA $rms$ current. In case where one horn is damaged and in order to keep the neutrino intensity intact without repairing that horn, the power supply unit (PSU) has to deliver pulses at the higher frequency of 16.66~Hz and 10.4~kA $rms$  to the remaining 3 horns. The amount of $rms$ current of the power supply is very important and therefore is necessary to design a solution that minimizes energy consumption. 

The PSU design has been studied to have a lifetime of more than 10 years assuming neutrino beam operation for 200 days per year continuously. The components of the PSU is foreseen to be placed as near to the horns as allowed with respect of the high level of radiation produced by the secondary particles emitted from the proton interactions with the target materiel. The steel and concrete shielding have been studied \cite{designreport}. Furthermore, the geometry and length of the strip-lines connecting the PSU with the horn has been optimized to keep their impedance at minimum.    

\begin{figure}[h]
\begin{center}
\includegraphics[width=.9\textwidth]{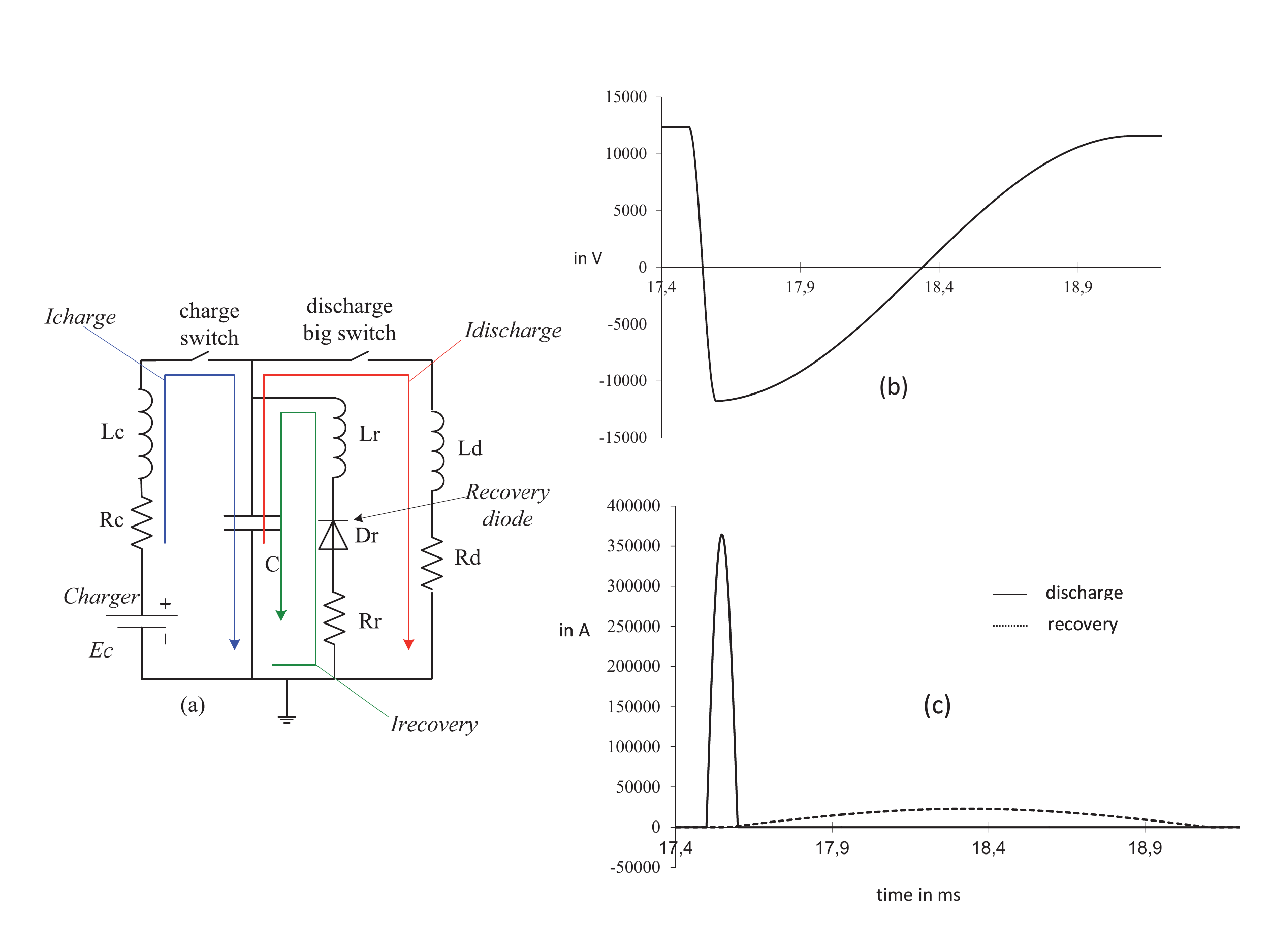}
\caption{\label{PSU_Horns_figure1} Principle of the power supply operation for a horn: (a) circuit-schematic: charging (blue), discharging (red) and recovery (dark-green) phases of PSU (b) capacitor-voltage vs. time in ms, (c) discharging and recovery current vs. time in ms.}
\end{center}
\end{figure}

\section{Principle of basic circuit}

The power supply design is based on the principle of charging a capacitor to +12~kV reference voltage and then discharging it into the horn using a big switch via a direct coupled electrical design thus providing the half-sinusoid waveform of 350~kA peak current necessary for the creation of the magnetic field in the horn. In detail and considering figure \ref{PSU_Horns_figure1},  when the switch of the continuous $E{\rm c}$ charger is closed, the capacitor $C$ is charged to +12~kV  during the first 17.5~ms through an oscillated circuit with parameters $R{\rm c},L{\rm c}$ and~$C$. At time T=17.5~ms, the big switch (responsible for the discharging) is closed and therefore the capacitor is discharged into the horn through an oscillated circuit with parameters $R{\rm d},L{\rm d}$ and~$C$. During the discharging cycle, the voltage of the capacitor becomes negative which allows the recovery diode $D{\rm r}$ to conduct current to a recovery coil $L{\rm r}$. Then, the voltage is inverted rapidly allowing the reduction of charging during the next cycle. This circuit also maintains high lifetime for the charging capacitor. 

The evolution of current and voltage during the charging, discharging and recovery cycles of the capacitor are governed by a second order differential equation of the corresponding resonant RLC circuit in the PSU. These equations are given in \cite{designreport7} and allow us to calculate, for each cycle, the final voltage $V_{\rm end}$, the value of the maximum current $I_{\rm peak}$ and the maximal slew rate $\rmd I/\rmd t$ of the capacitor. In summary, the waveforms of the capacitor current and voltage during discharging and recovery stages are given by the following formulas:

\begin{equation}
\frac{\displaystyle{\rmd^{2}V}}{\displaystyle{\rmd t}} + 2m\omega \frac{\displaystyle{\rmd V}}{\displaystyle{\rmd t}} + \omega^{2}V = 0, ~I(t) = -C\frac{\displaystyle{\rmd V}}{\displaystyle{\rmd t}} \label{eq:0.1}
\end{equation}

For $m < 1$ then:

\begin{equation}
V(t) = V_{\rm 0} \rme^{-m\omega t}(\cos \omega t + m\sin \omega t) \label{eq:0.2}
\end{equation}

\begin{equation}
I(t) = V_{\rm 0}C\omega \rme^{-m\omega t} \sin \omega t \label{eq:0.3}
\end{equation}

\begin{equation}
\frac{\displaystyle{\rmd I(t)}}{\displaystyle{\rmd t}} =  V_{\rm 0}C\omega^{2} \rme^{-m\omega t} (\cos \omega t - m\sin \omega t) \label{eq:0.4}
\end{equation}
where  $V_{\rm 0}$ is the initial capacitor voltage, $\omega = 1/\sqrt{LC}$ is the angular frequency and $m = \frac{\displaystyle{R}}{\displaystyle{2}} \sqrt{\frac{\displaystyle{L}}{\displaystyle{C}}}$ is the damping value.  With a near zero value of $m$, the waveform of the capacitor current is practically a perfect half sinusoid. 

\begin{table}[h]
\caption{Simulation results for the electrical constraints.\label{tab:station}}
\begin{indented}
\item[]\begin{tabular}{@{}*4{l}}
\br
$C$ = 1000 $\mu$F          & Charging             & Discharging                          &  Recovery    \\ 
$E{\rm c}$ = 10980 V       & $L{\rm c}$ = 25 mH   & $L{\rm d}$ = 1.041 $\mu$H              &  $L{\rm r}$ = 0.25 mH   \\        
$I_{\rm peak}$ = 350 kA    & $R{\rm c}$ = 0.475 $\Omega$  & $R{\rm d}$ = 0.89 m$\Omega$    &  $R{\rm r}$ = 5.94 m$\Omega$ \\
\mr
m                          & 0.0475               & 0.013                                  &  0.005724    \\

$\omega$ (rd/s)            & 200                  & 30990                                  &  1928        \\ 

${\rmd I(t = 0)}/{\rmd t}$~(A/$\mu$s) & 0.014 & 10846                                  &  43          \\

$V_{\rm 0}$ (V)            & {\bf 10616}                & 11293                                  &  -10816      \\ 

$V_{\rm end}$ (V)          & 11293                & -10816                                 &  10616       \\ 

$I_{\rm peak}$ (A)         & 68                   & 350~k                                  &  {\bf 21.6~k}       \\ 
\br
\end{tabular}
\end{indented}
\end{table}

In the hypothesis that a unique power supply module is used to provide 350~kA peak current, the values for critical currents and voltages are calculated and presented in table \ref{tab:station}. 

The value of $\rmd I(t)/\rmd t$ in the discharging stage is too high to be supported by the recently known power thyristors with maximum specification of 1.5 kA/$\mu$s. In addition, the peak current of the recovery stage is very high and as result it is not possible for the manufacturers to provide such a coil. Therefore, a modular approach is followed to reduce the electrical constraints on the various components and explained in the sections. 

\begin{figure}[t]
\begin{center}
\includegraphics[width=10.cm]{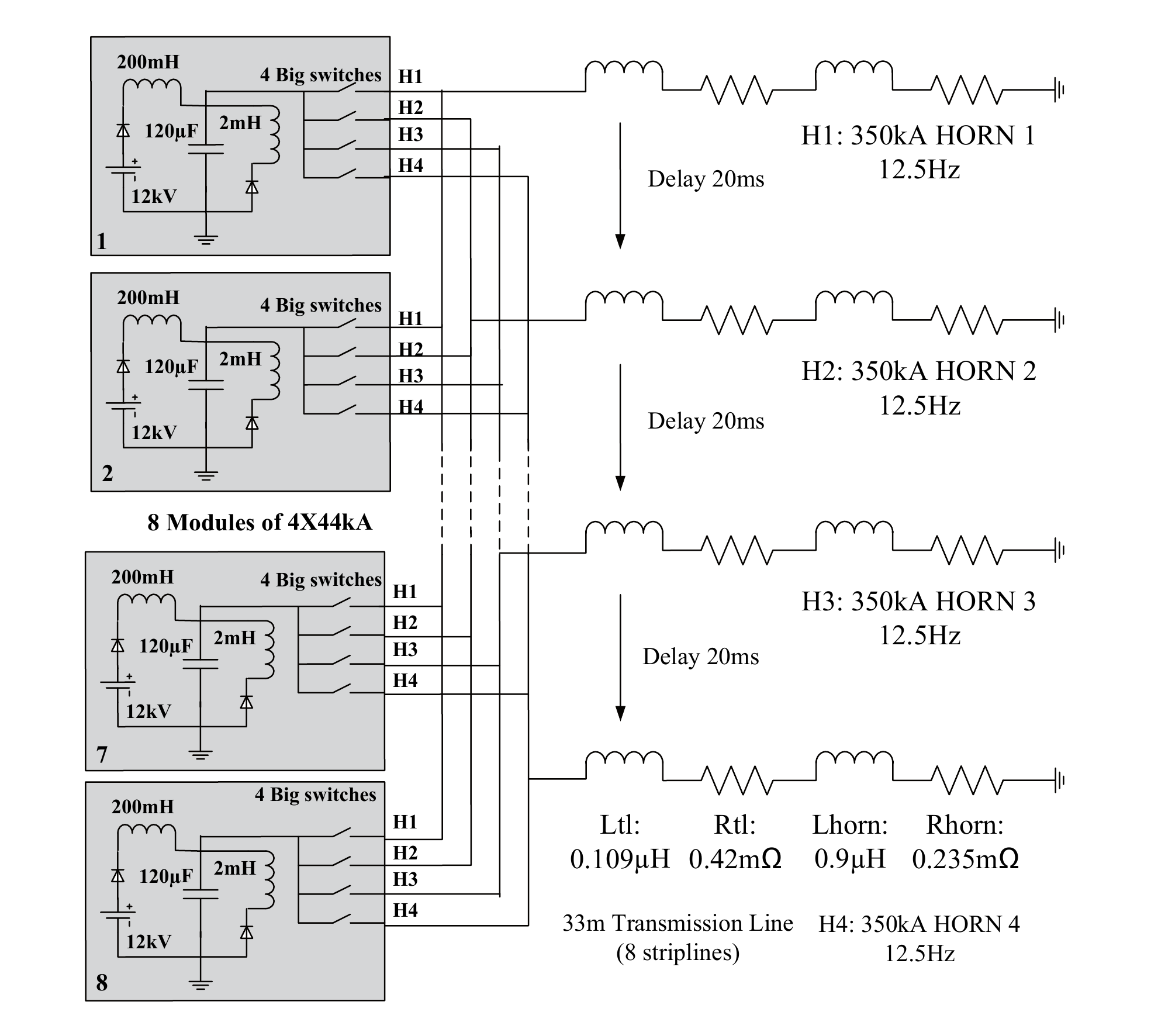}
\caption{\label{PSU_Horns_figure2} Principle of the modular architecture in order to supply 350~kA to each horn by using 8 modules. Each module provides four outputs of 44~kA at 12.5~Hz.}
\end{center}
\end{figure}

\begin{figure}[h]
\begin{center}
\includegraphics[width=12.cm]{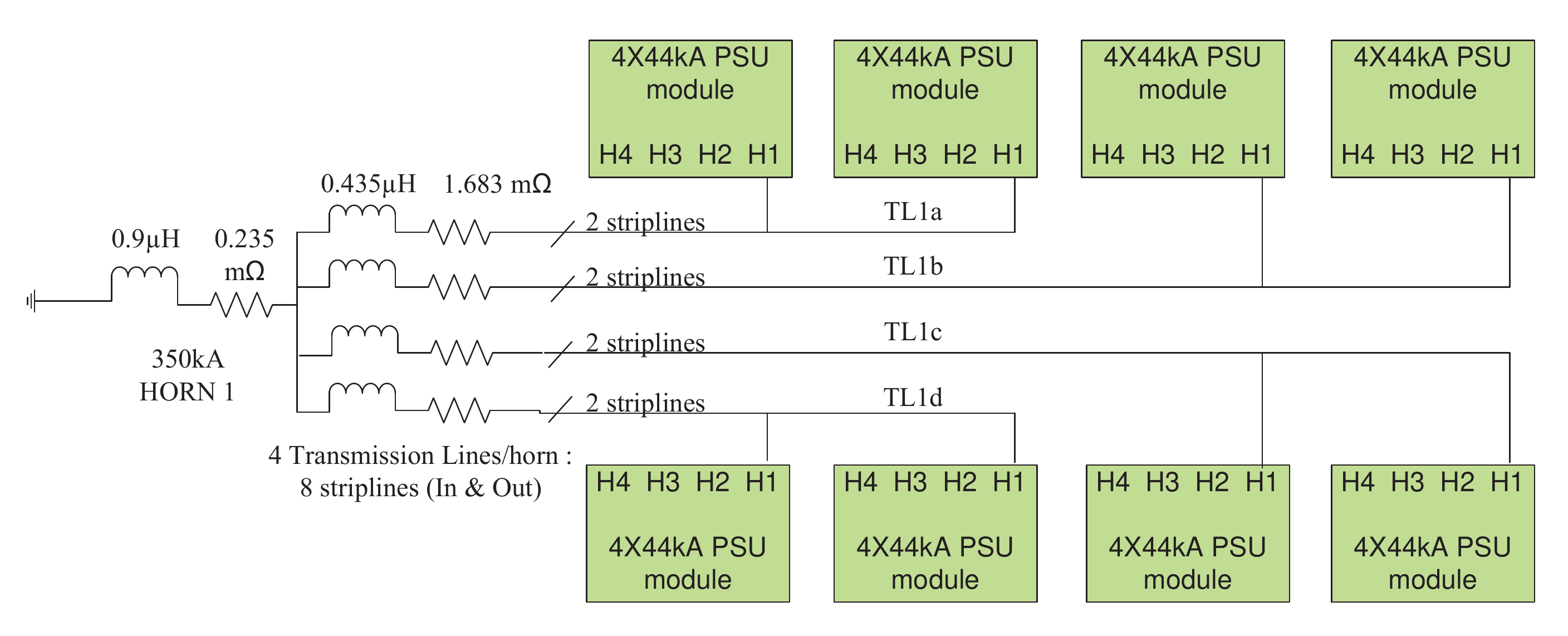}
\caption{\label{PSU_Horns_figure3} Layout of the transmission and strip lines for one horn providing 350~kA at 12.5~Hz.}
\end{center}
\end{figure}

\section{Implementation of the solution}

A system with 8 modules (figure \ref{PSU_Horns_figure2}) have been chosen in order to reduce the length of the transmission lines and the price of the PSU. Each module provides four sequential current outputs of 44~kA and every 20~ms (50~Hz) to the four horns and referred as 4$\times$44~kA module.

For each one of the 8 modules, the capacitor charging and recovery circuits operate at 50~Hz while its discharging circuit on each horn by a big switch operates at 12.5~Hz through one of its four 44~kA outputs. Thus a peak current of about 350~kA is provided  through strip-lines to each horn every 80~ms (12.5~Hz) or every 60~ms (16.66~Hz) depending on the number of horns used: four for normal operation or three in case of a damaged horn. Furthermore, two modules are interconnected  by two strip-lines to the same transmission line with  $R{\rm tl}$ = 1.683 m$\Omega$ and $L{\rm tl}$ = 435~nH  (figure \ref{PSU_Horns_figure3}). In order to limit the energy consumption,  investigations have been done to reduce the resistivity and inductance by studying a transmission line based on large aluminium strip-lines \cite{designreport1}. This allows us to obtain a small resistivity of 51~$\mu \Omega$/m and 13.2~nH/m for 2 plates (0.6~m high times 1~cm width) spaced by 1~cm. The total length of a transmission line does not exceed 33~m.

\begin{figure}[t]
\begin{center}
\includegraphics[width=11.cm]{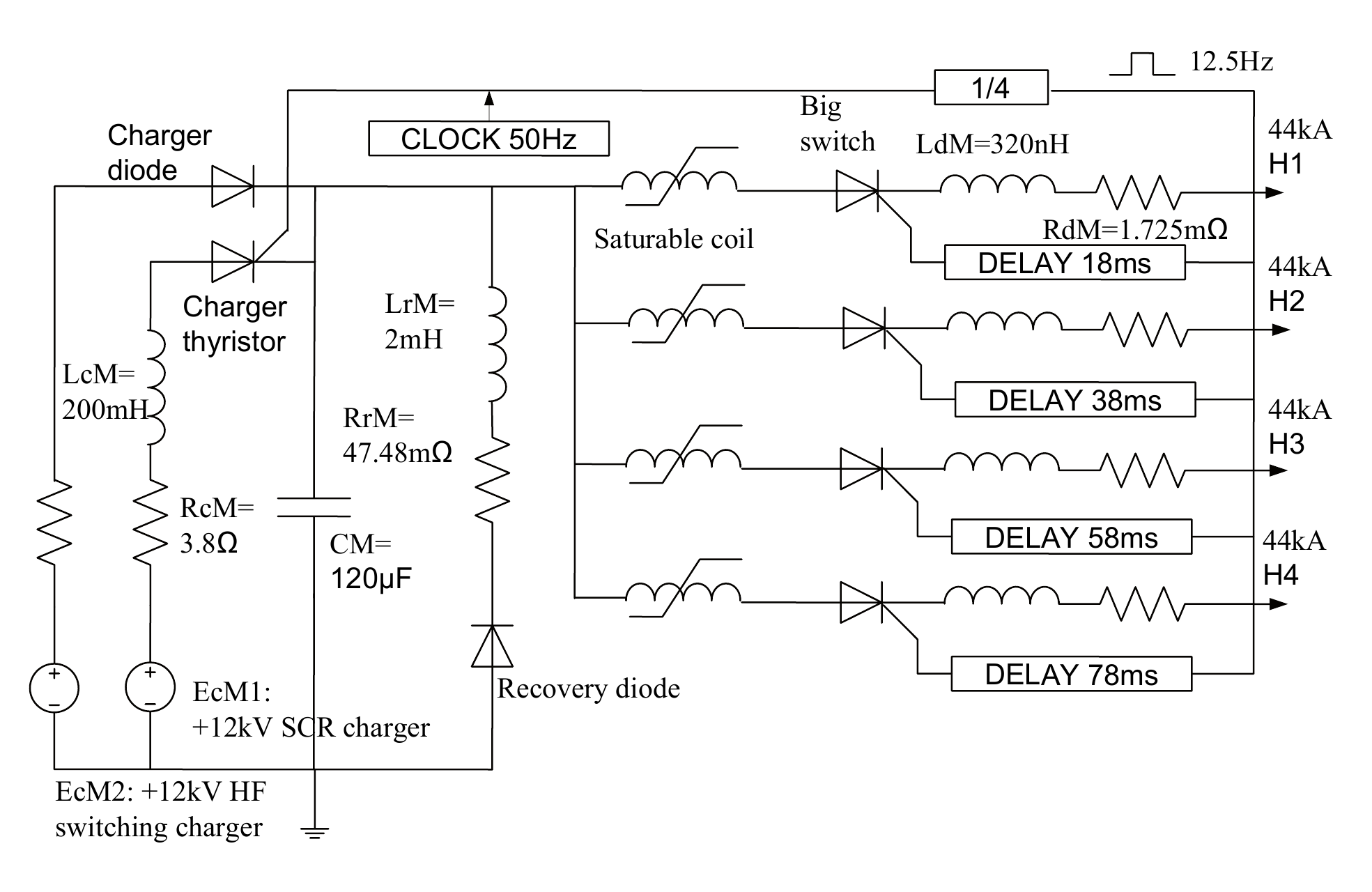}
\caption{\label{PSU_Horns_figure4} Functional principle of a 4$\times$44 kA PSU module.}
\end{center}
\end{figure}

\subsection{ 4$\times$44 kA PSU module description}

The neutrino beam will be functioning for 10 years, 200 days per year or 1.7 10$^8$~s, thus the PSU components should have a cycle lifetime of the order of few billions cycles. In order a safety margin to exist,  the components have been studied on the basis of a lifetime of 13 billions of cycles for the charging and recovery stages, and 3.25 billions of cycles for the big switches of discharging stages.  The values of $I_{\rm FRMS}$ (maximum RMS current at on-state), $I_{\rm peak}$ and $Q$ (the amount of electrical charge) in the capacitor during the charging, recovery and discharging stages have been estimated by simulations and are given in table~\ref{tab:station2}. They match the values of PSU electrical constraints (table \ref{tab:station}) with scaling 1 to 8, thus confirming that 8 of 4$\times$44~kA modules are needed for the four-horn system. The specific values of equivalent inductance and resistance of each stage will be explained later and are depending on the electrical characteristics of the components. In figure \ref{PSU_Horns_figure4}, the schematics of the module and its components are shown. 

\begin{table}[h]
\caption{Current constraints for each stage of the 4$\times$44~kA module: results of PSIM simulations \cite{PSIM}.\label{tab:station2}}
\begin{indented}
\item[]\begin{tabular}{@{}*{7}{l}} 
\br
$I$                      & $\rmd I(t = 0)/\rmd t$ & $I_{\rm peak}$   & \centre{3} {$I_{\rm FRMS}$}    & $Q$       \\ 
 
constraints stages       & (A/$\mu$s)              & (A)             & \centre{3} {(A)           }    & (C)      \\    
\ns
& & & \crule{3} & \\ 
$E{\rm c}$ = +12 kV       & Nominal       &        & 50 Hz           & 16.66 Hz     & 12.5 Hz              & Nominal \\
\mr
Charging                  & 0.002         & 9.56   & 5.9             & ---            & ---                    & 0.09    \\ 

Recovery                  & 5.88          & 2864   & 562             & ---            & ---                    & 2.81    \\

Discharging               & 1461          & 45.56  & 2278            & 1315         & 1139                 & 2.9     \\
\br
\end{tabular}
\end{indented}
\end{table}

The power delivered by the capacitor charger attains 70.8~kW rms per module, thus 566~kW rms for the total PSU. It represents only 3\% of the current amount discharged on the horn, therefore the recovery energy efficiency is very high at 97\%.

\begin{figure}[t]
\begin{minipage}{17pc}
\begin{center}
\includegraphics[width=17pc]{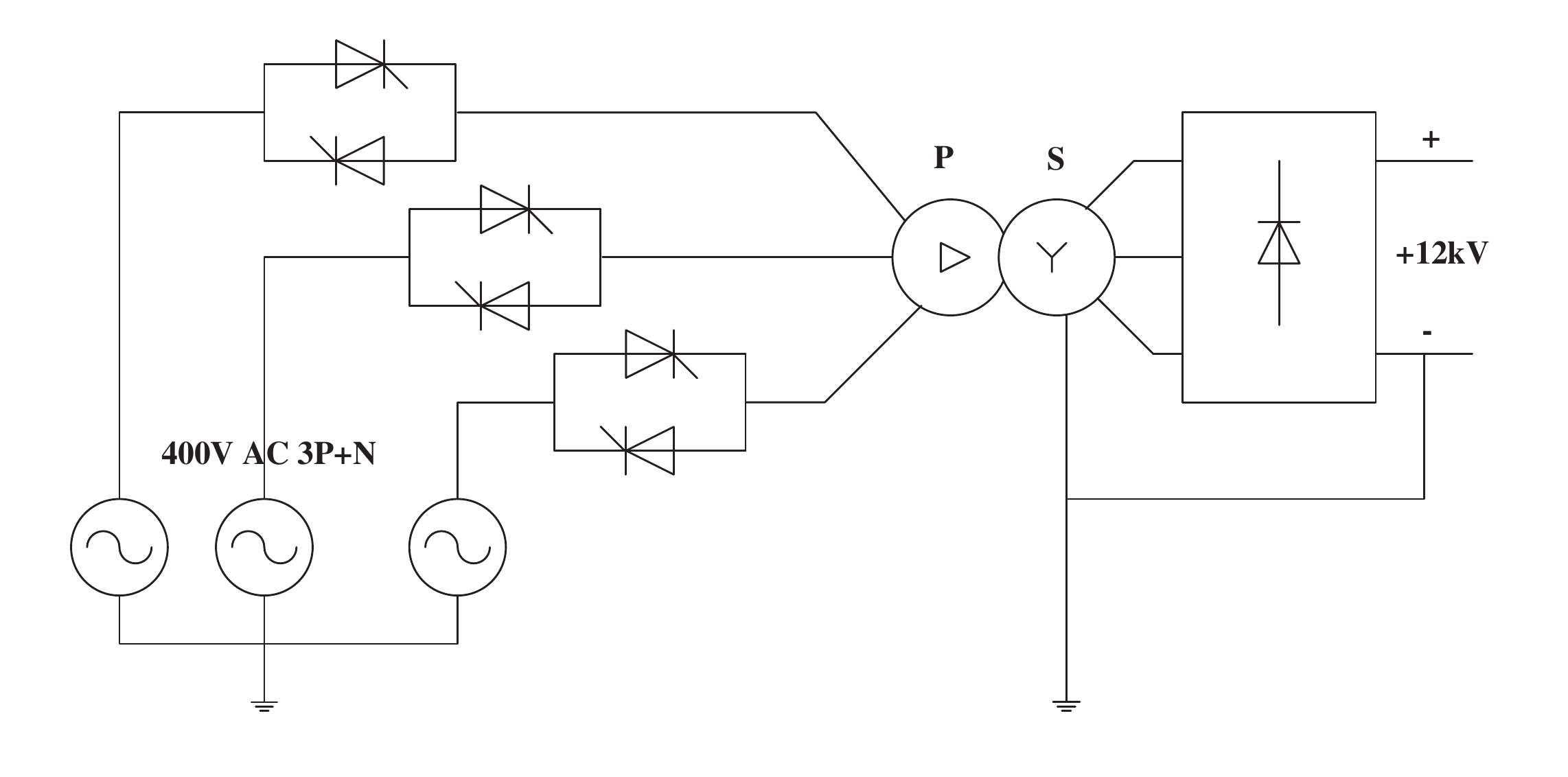}
\caption{\label{PSU_Horns_figure5} Silicon Controlled Rectifier (SCR) 3 Phases AC-DC converter charger EcM1.}
\end{center}
\end{minipage}\hspace{2pc}
\begin{minipage}{17pc}
\begin{center}
\includegraphics[width=17pc]{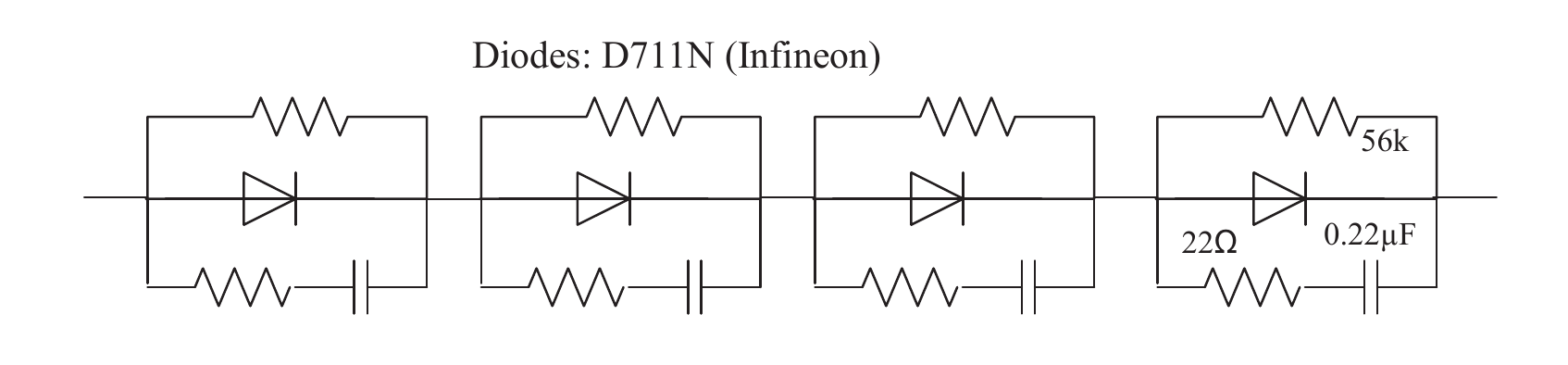}
\caption{\label{PSU_Horns_figure6} Recovery diode diagram, diodes mounted with resistance sharing protection ($R$~=~56~k$\Omega$) and snubber circuit ($R$~=~22~$\Omega$, $C$~=~0.22~$\mu$F).}
\end{center}
\end{minipage}
\end{figure}

\subsubsection{ Electrical charger module}

Studies have been performed for a capacitor charger to provide the high power of 70~kW at a high +12~kV voltage in 18~ms maximum with high reliability, high band pass and low electromagnetic interference (EMI) perturbations. A hybrid solution is proposed (figure \ref{PSU_Horns_figure4}): it is based on the association of a high reliability basic Silicon Controlled Rectifier (SCR) 3-phases converter AC-DC solution EcM1 (figure \ref{PSU_Horns_figure5}) which provides 90\% of power in 16.5~ms and a high frequency switching charger EcM2 (at 25~KHz) which increases the bandwidth of the charger, offers a very good reactivity to fast 400~VAC supply variations ($\pm$1\% in 40~ms max) and guarantees fast stabilization at $\pm$0.5\% in 18~ms maximum.

\begin{table}[h]
\caption{Recovery function: comparison between maximum specifications and calculated values from simulations for the diode ($V_{\rm RRM}$: maximum repetitive peak reverse voltage; $Q_{\rm RR}$: maximum reverse recovery charge; $I_{\rm R}$: maximum reverse current).\label{tab:station3}}
\begin{indented}
\item[]\begin{tabular}{@{}llll}
\br
Parameter       & $V_{\rm RRM}$   & $Q_{\rm RR}$   & $\rmd I_{\rm R}(t = 0)/\rmd t$ \\ 
                & (kV)            & (mC)           & (A/$\mu$s)      \\       
\mr
D711N diode     & 7               & 5.5            & 10              \\

Simulations     & 3.5             & 3.2            & 5.88            \\ 
\br
\end{tabular}
\end{indented}
\end{table}

\subsubsection{Recovery stage module}

The big 2~mH LrM inductactor (figure \ref{PSU_Horns_figure4}) will be made of an iron coil with a low saturation of 3\% at maximum current. It is made in molded winding and guarantees high isolation of 25~kV between wires and ground. With a low resistivity of 43.75~m$\Omega$ the copper and iron losses are 13.8~kW, and 3.2~kW respectively. The total 17~kW will be dissipated by water cooling methods. 

Press pack D711N-Infineon \cite{components} diodes have been chosen for the recovery diode function (figure \ref{PSU_Horns_figure6}). In table \ref{tab:station3}, the reverse electrical constraints per diode are shown. The calculated values are two times below their specifications given in their datasheets \cite{components} which indicates high reliability.

The total power $P_{\rm diodes}$ dissipated by the 4 diodes is the sum of conduction and commutation thermal losses with the specific values of the D711N-Infineon diode given in their datasheet \cite{components}. Applying the average on-state current $I_{\rm FAV} = F\times Q$, the maximum threshold voltage $V_{\rm T0max}$ = 0.84~V, the slope resistance $R_{\rm T}$ = 0.87~m$\Omega$, operation frequency F = 50~Hz, to the following formula \ref{eq:1}, the power is :

\begin{equation}
 P_{\rm diodes} = 4(V_{\rm T0max}I_{\rm FAV} + R_{\rm T}I_{\rm FRMS}^2 + V_{\rm RRM}FQ_{\rm RR}) = 4.7~{\rm kW} \label{eq:1}
\end{equation}

In order to dissipate so high power, the diodes will be installed on an aluminium high thermal exchanger cooled by water.

\begin{figure}[t]
\begin{center}
\includegraphics[width=10.cm]{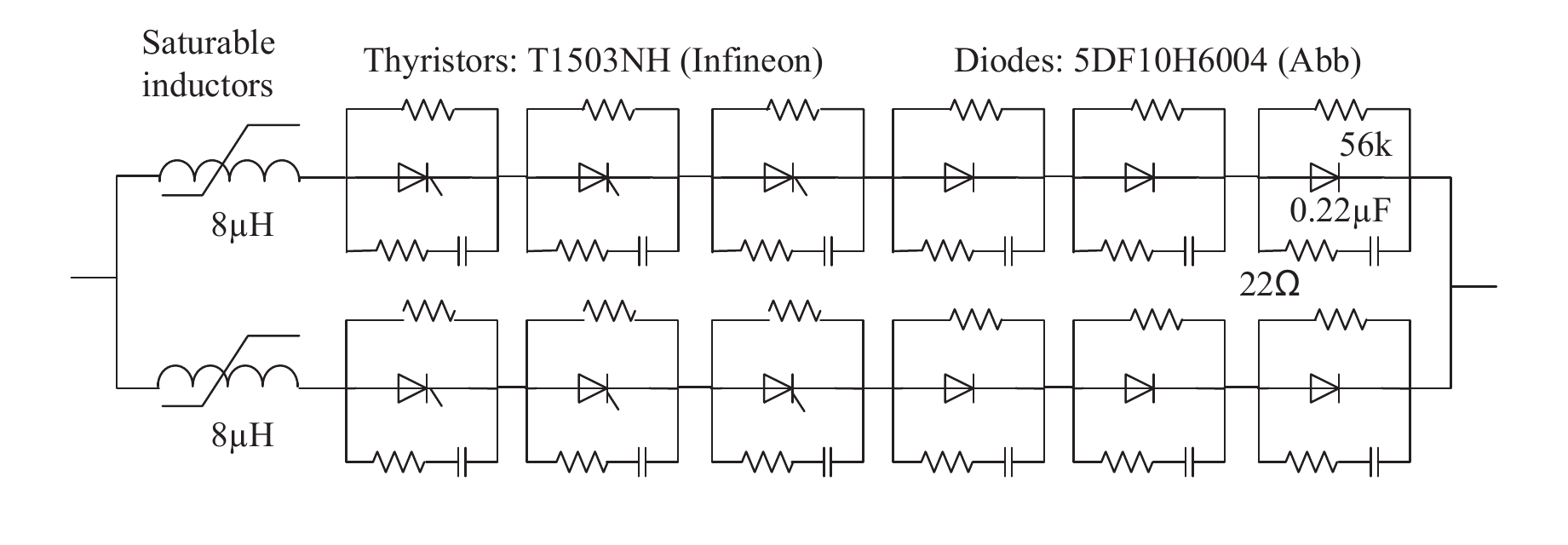}
\caption{\label{PSU_Horns_figure7} Discharge big switch diagram, thyristors and diodes mounted with resistance sharing protection ($R$~=~56~k$\Omega$) and snubber circuit ($R$~=~22~$\Omega$, $C$~=~0.22~$\mu$F).}
\end{center}
\end{figure}

\subsubsection{Discharging stage module} 

The solution for the discharging big switch (figure \ref{PSU_Horns_figure4}) is shown in figure \ref{PSU_Horns_figure7}. It is assembled with 2 sections of 3 press-pack thyristors (T1503-Infineon) serially placed with 3 press-pack fast recovery diodes (5DF10H6004-Abb) \cite{components}. The thyristors allow to block the very high direct +12~kV  voltage ($V_{\rm DRM}$) by closing the switch rapidly (1-2~$\mu$s). Fast diodes are rapidly blocking negative current. Saturable coils are limiting $\rmd I(t)/\rmd t$ during critical switching on and off states, to 150 A/$\mu$s in each section. 

\begin{table}[h]
\caption{Comparison between calculated and maximum specification values of critical parameters for components at discharging stage ($I_{\rm FSM}-I_{\rm TSM}$: maximum surge forward current; $I_{\rm TRMS}-I_{\rm FRMS}$: maximum RMS forward current; $V_{\rm DRM}$: maximum direct voltage; $\rmd Icr/\rmd t$: maximum critical rate of rise of on-state current repetitive).\label{tab:station4}}
{\footnotesize
\begin{tabular}{@{}*{11}{l}}
\br
          & \centre{5}{Thyristor T1503 Infineon} & \centre{5}{Diode 5DF10H6004 Abb}\\
\ns
          &    \crule{5}                         &  \crule{5} \\
Par.      & $I_{\rm TSM}$ & $I_{\rm TRMS}$ & $V_{\rm DRM}$ & ${\rmd Icr(0)}/{\rmd t}$&${\rmd I(0)}/{\rmd t}$ & $I_{\rm FSM}$ & $I_{\rm FRMS}$ & $V_{\rm RRM}$ & $Q_{\rm RR}$ & ${\rmd I_{\rm R}(0)}/{\rmd t}$ \\ 
             & (kA)     & (kA)    & (kV)    & (kA/$\mu$s)   & (kA/$\mu$s) & (kA)  & (kA)    & (kV)  & (mC)  & (A/$\mu$s) \\   
\mr
Spec.         & 40     & 3.8   & 7.5   & 0.3     & 5     & 18   & 1.7  & 6   & 6   & 0.3 \\  

Calc.         & 22.8   & 0.66  & 4     & 0.15    & 0.75  & 22.8 & 0.66 & 4.5 & 4.1 & 0.15 \\
\br
\end{tabular}}
\end{table}

The calculated values of critical parameters applied on the thyristors and diodes are compared  to maximum specification values listed in table \ref{tab:station4}. As result, the calculated values are well below the maximum values (excluded $I_{\rm FSM}$). 

Only conduction power $P_{\rm thyristors}$ is dissipated by the 6 thyristors and is calculated from the formula: 

\begin{equation}
P_{\rm thyristors} = 6(V_{\rm T0max}I_{\rm TAV}+R_{\rm T}I^{2}_{\rm TRMS}) = 1.5~{\rm kW}
\end{equation}
with the specific values of the thyristor T1503-Infineon (i.e. $I_{\rm TAV} = F \times Q$, {$V_{\rm T0max}$ = 1.24~V, $R_{\rm T}$ = 0.44~m$\Omega$, $F$ = 16.66~Hz) given in their datasheet \cite{components}. The thyristors will be mounted on aluminium exchangers with water cooling.

Conduction and commutation thermal losses are dissipated by the 6 diodes and calculated from the equation \ref{eq:1} modified for 6 parts with the specific values of the diode 5DF10H6004-Abb (i.e. $V_{\rm T0max}$ = 1.5~V, $R_{\rm T}$ = 0.6~m$\Omega$, $F$ = 16.66~Hz) given in their datasheet \cite{components}. The maximum power dissipated is $P_{\rm diodes}$ = 3.84~kW. Again, the diodes will be mounted on aluminium high-thermal exchangers with water cooling.

As result, the total maximum power dissipated is 5.3~kW when three horns are running, each at 16.66~Hz, in order to keep the particle intensity intact (one out of four horns has failed). In nominal mode operation when four horns are running, each running at 12.5~Hz, the power dissipation will be less and equal to 4~kW.

\begin{table}[h]
\caption{Electrical resistivity and inductance of components of the 4$\times$44~kA module.\label{tab:station5}}
\begin{indented}
\item[]\begin{tabular}{@{}*{7}{l}}
\br  
                & \centre{2}{Charge} &\centre{2}{Discharge} &\centre{2}{Recovery}\\ 
\ns
                & \crule{2} & \crule{2} & \crule{2}   \\   
                & LcM       & RcM      & LdM       & RdM      & LrM      & RmM     \\ 
                & (H)    & ($\Omega$)   & (nH)        & (m$\Omega$)  & (mH)       & (m$\Omega$) \\
\mr
$L$ charge coil    & 0.2       &   ---       &    ---       &     ---     &    ---      &    ---     \\

$R$ charge coil    & ---          & 0.8      &      ---     &    ---      &    ---      &    ---     \\

$R$ charger Ec     &    ---       & 3        &    ---       &     ---     &     ---     &   ---      \\       

$L$ capa 120$\mu$F    &    ---       &    ---      & 50        &   ---       &    ---      &   ---      \\                 

$R$ capa 120$\mu$F    &    ---       &    ---      &   ---        & 0.2      &   ---       &   ---      \\       

$L$ saturable coil&      ---     &     ---     & 120       &    ---      &   ---       &    ---     \\                 

$R$ saturable coil&     ---      &    ---      &   ---        & 0.2      &   ---       &   ---      \\                 

$R$ 3diodes + 3 thyristors & --- &    ---      &   ---        & 3.12     &  ---        &   ---      \\ 

$L$ 3diodes + 3 thyristors & --- &    ---      & 300       &    ---      &   ---       &   ---      \\                 

$R$ recovery diode&      ---     &    ---      &   ---        &    ---      &   ---       & 3.48    \\

$L$ recovery coil &     ---      &     ---     &   ---        &    ---      & 2        &   ---      \\ 

$R$ recovery coil &     ---      &     ---     &    ---       &     ---     &   ---       & 44      \\ 
\mr
{\bf TOTAL}     & {\bf 0.2} & {\bf 3.8}& {\bf 260} & {\bf 1.86}& {\bf 2} & {\bf 47.48}\\
\br
\end{tabular}
\end{indented}
\end{table}

\begin{table}[h]
\caption{Summary of thermal dissipations from components of the 4$\times$44~kA module.\label{tab:station6}}
{\footnotesize
\begin{tabular}{@{}*{7}{l}}

\br  
4$\times$44 kA PSU      & \centre{6} {Module dissipated power (W)}\\ 
\mr
                & \centre{3}{water cooling}                  & \centre{3}{air cooling} \\
                & \centre{3}{conductivity water 1-10 $\mu$S} & \centre{3}{}            \\
\ns
                & \crule{3}                                  & \crule{3}               \\
                & Charge                     & Discharge& Recovery  & Charge   & Discharge& Recovery   \\ 
\mr
Charger +12 kV - 5.9 Arms & 7100            &    ---       &    ---        &    ---       &   ---        &      ---        \\

Capacitors bench 120 $\mu$F&    ---     &    ---       &  ---          &  ---         & 992         &   ---           \\

Recovery coil 2 mH     &       ---       &   ---        & 17000     &    ---       &   ---        &    ---          \\

Recovery diode        &        ---     &    ---       & 4700      &     ---      &    ---       &     ---         \\

4 big switches        &      ---       & 4000$\times$4   &   ---         &   ---        &    ---       &     ---         \\  

Saturable coils       &     ---       &   ---     &  ---     &    ---     & 200$\times$2$\times$4  &    ---          \\  
                & \crule{3}                                  & \crule{3}               \\
{\bf TOTAL per MODULE}& \centre{3}{\bf 45 kW}      & \centre{3}{\bf 2.6 kW} \\  
\br
\end{tabular}}
\end{table}

\begin{figure}[t]
\begin{center}
\includegraphics[width=8.cm]{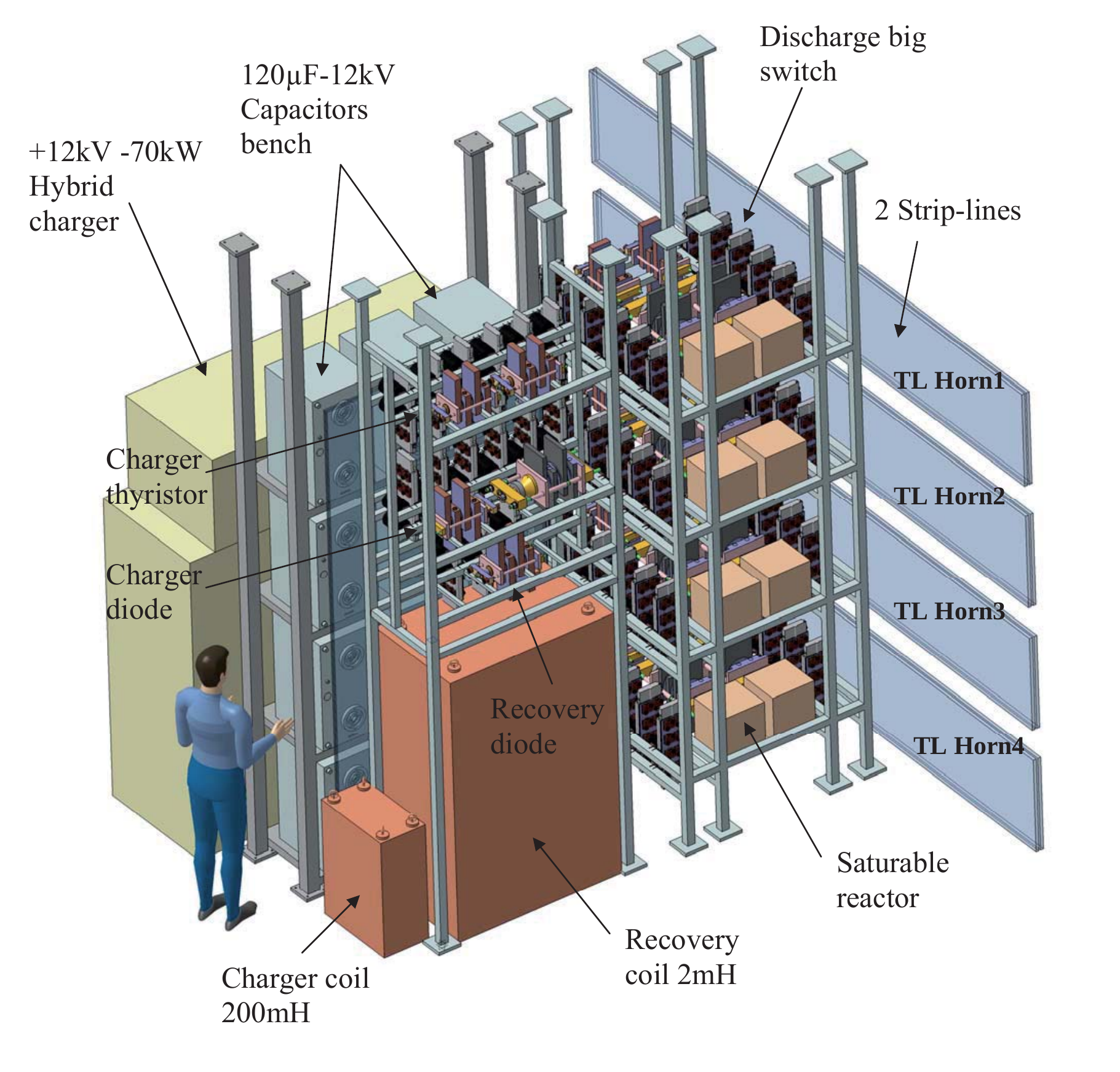}
\caption{\label{PSU_Horns_figure8} View of 4$\times$44 kA module.}
\end{center}
\end{figure}

\begin{table}[h]
\caption{Summary of component dimensions and weights of the 4$\times$44~kA module.\label{tab:station7}}
\begin{indented}
\item[]\begin{tabular}{@{}lll}
\br
PSU 4$\times$44~kA module        & Dimensions,         & Weight,   \\ 
                         & width$\times$high$\times$depth~m & ~kg        \\ 
\mr
Discharge big switch     & 1.35$\times$3.2$\times$1.55       & 400       \\ 

Recovery coil 2~mH       & 1.1$\times$1.7$\times$0.7         & 2200      \\ 

Recovery diode           & 1.4$\times$0.525$\times$0.72      & 100       \\ 

Hybrid charger 12~kV -70~kW & 2.1$\times$2.53$\times$0.8        & 2400      \\ 

120~µF +-12~kV capacitors bench & 1.73$\times$3.2$\times$0.63  & 1700      \\ 

Charger coil 200~mH      & 0.3$\times$0.75$\times$0.55       & 190       \\ 

Charger thyristor        & 1.4$\times$0.525$\times$0.72      & 100       \\ 

Charger diode            & 1.4$\times$0.525$\times$0.72      & 50        \\ 

3 Mechanicals supports   &      ---           & 500       \\ 
\mr
{\bf TOTAL module}       & {\bf 3.43$\times$3.525$\times$2.41}&{\bf 8840} \\ 
\br
\end{tabular}
\end{indented}
\end{table}

\subsection{Implementation of the 4$\times$44~kA PSU module}

A 4$\times$44~kA module is shown in figure \ref{PSU_Horns_figure8}. In order to reduce the resistivity and inductance in the discharging and recovery circuits: a) the +12 kV and 20~$\mu$F unit capacitors must be connected on two large strip-lines spaced by 1~cm, b) the discharging switches must be placed as near as possible to the horn strip-lines and the output of the capacitors, and c) the recovery components (coil, diode) must be directly interconnected at the output of the capacitors. Also, the strip-lines of return currents must be placed close to the press-pack components to limit the discharging switch inductance. That works by adding four return conductors around the switch stack \cite{designreport3}. In table \ref{tab:station5}, the values of resistivity and inductance of components of the PSU modules are given. In table \ref{tab:station6}, the summary of the dissipations of several module components are given. The 95\% of the power will be dissipated by water cooling methods.

The dimensions and weight of different parts of the module are given in table \ref{tab:station7}. For the investigated Super Beam application, a 5~ton crane will also be sufficient to manipulate the 1.75~ton of the four discharge big switches. Finally, each PSU module occupies 8.3~m$^2$ (figure \ref{PSU_Horns_figure8}) of space. The ground mechanical constraints are 1~ton/m$^2$ in average and they can attain a maximum of 20~ton/m$^2$ at the contact between ground and supports feets. 

\begin{figure}[h]
\centering
\includegraphics[width=10cm]{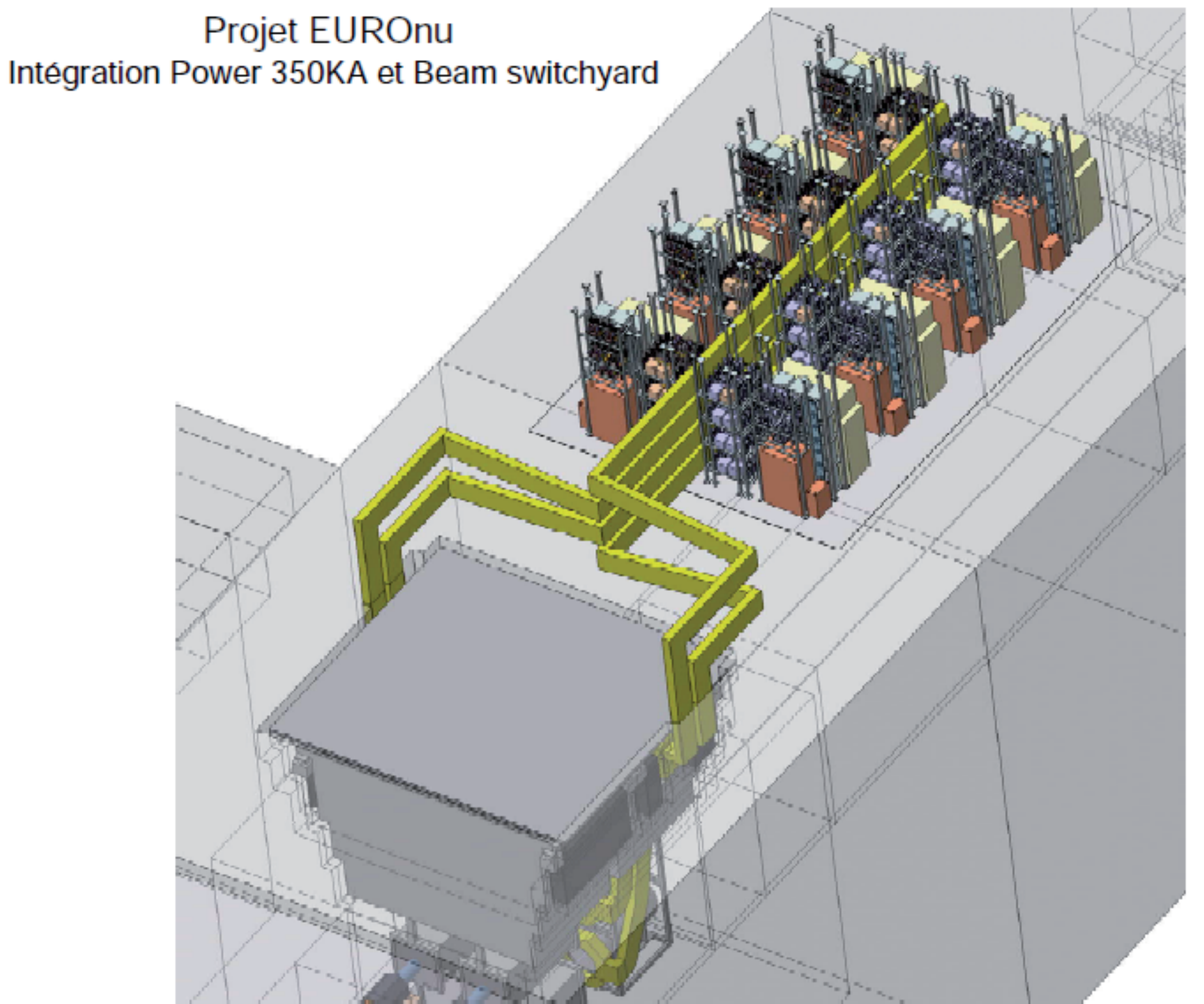}
\caption{\label{psupply} 3D layout of the PSU placed on top of the target station and decay tunnel.}
\end{figure}

\begin{figure}[h]
\centering
\includegraphics[width=10cm]{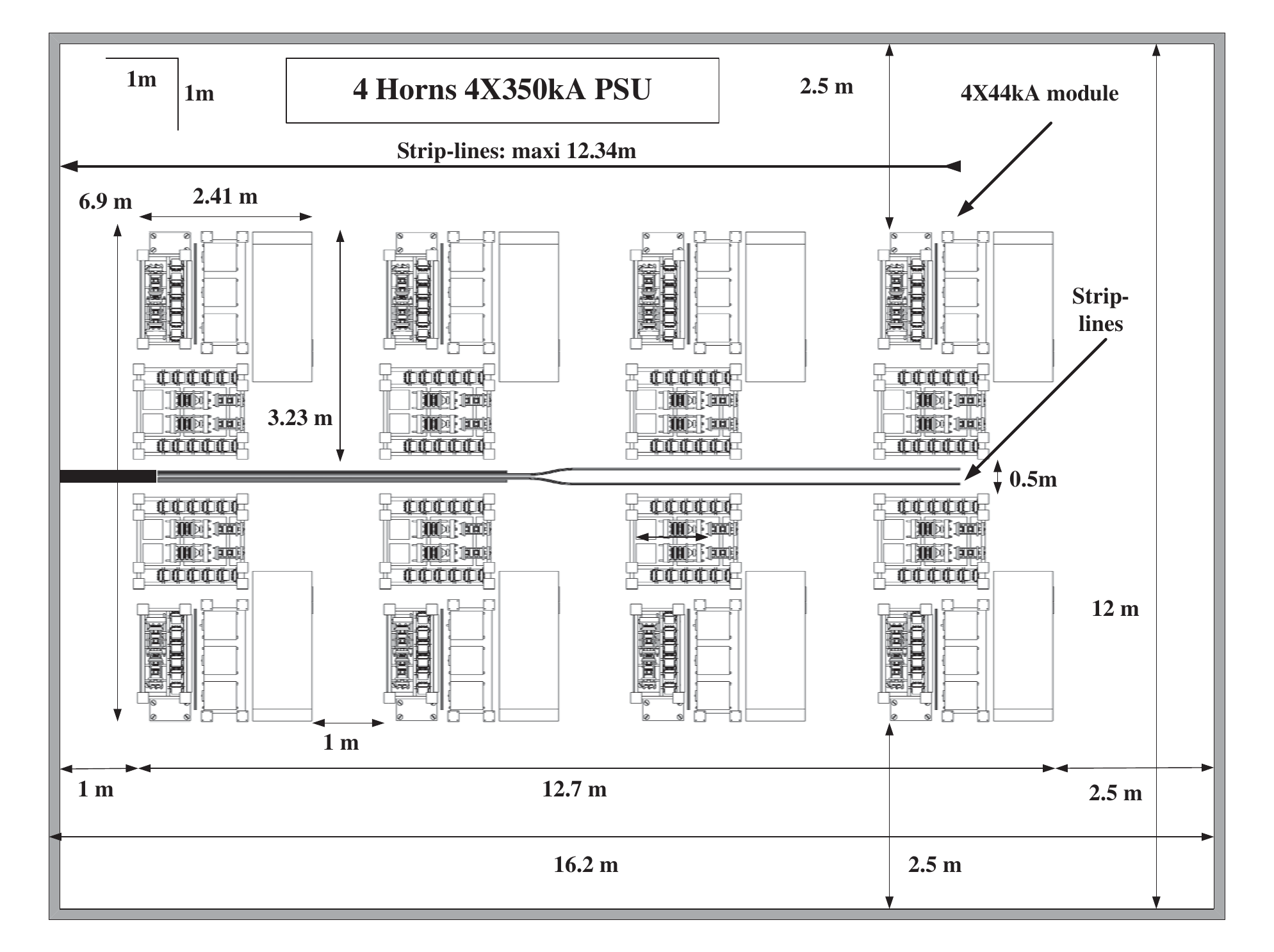}
\caption{\label{PSU_Horns_figure9} Top view of the PSU room.}
\end{figure}

\section{Complete power supply unit}

For the whole PSU, the 8 modules of 4$\times$44~kA  will be installed along the four vertical transmission lines on a total area of 194~m$^2$ (length 16.2~m $\times$ width 12~m, figure \ref{psupply} and  \ref{PSU_Horns_figure9}). Depending on future R\&D programs for the chargers and their final dimensions, it might be possible to rearrange the current configuration by installing the chargers along the peripheral of the room: that will reduce the total length of strip-lines to 30~m. A 5~ton crane is enough to move the large and robust frames that support the module components to a near maintenance zone area.

The civil engineering should take into account that the maximum height of the room has to be at least 4~m. Regardless of seismic constraints, PSU components supports should be attached on metallic robust transversal beams.

\begin{table}[b]
\caption{Summary of total PSU thermal dissipations.\label{tab:station8}}
{\footnotesize
\begin{tabular}{@{}*{6}{l}}
\br
                  &                     & \centre{4}{Total dissipated power (W)}             \\ 
\mr
                  &                     & \centre{2}{Water cooling}& \centre{2}{Gas cooling}\\
\ns
                  &                     & \crule{2}                & \crule{2} \\
                  &                     & No contaminate            & Contaminate & Air       &  Helium    \\
                  &                     & circuit                   & circuit     &           &        \\                  
\mr \ns
4 Horns            & 32 strip lines     &  ---                 &   ---           & 1750$\times$32   &   ---    \\
PSU room          & (L = 14~m)          &                      &             &  = 56~kW     &      \\
\ns \mr \ns
                  & Components          & 360~kW               &      ---        & 21~kW      &    ---   \\
\ns \mr \ns
Intermediate      & 32 strip lines      &          ---         &  1250$\times$32           &    &      \\
Room              & (L = 10~m)          &                       &  = 40~kW            &      &      \\
\ns \mr \ns
Vessel            & 32 strip lines      &        ---             &     ---         &     ---       & 1140$\times$32     \\
Room              & (L = 9.1~m)         &                          &             &           &    = 36.4~kW   \\
\ns \mr \ns
                  & 4 Horns             &         ---                   & 19000$\times$4     &    ---        &    ---   \\
                  & (Joule effect)      &                           &    = 76~kW     &           &      \\ 
\ns \mr \ns
                  & 4 Horns             &        ---                    & 23000$\times$4     &    ---        &  ---     \\
                  & (Deposition)        &                         &  = 92~kW        &           &      \\ 
\ns \mr \ns
                  & 4 Targets           &        ---                  &    ---          &    ---        & 85000$\times$4    \\
                  &                     &                         &             &           &  = 340~kW     \\
\mr 
{\bf TOTAL}       &                     &{\bf 360~kW}            &{\bf 208~kW}&{\bf 77~kW}&{\bf 380~kW}\\                   
\br
\end{tabular}}
\end{table}

\subsection{Safety}

Human and building safety has to be taken into account. PSU components (strip-lines, modules) have to be protected from human access during operation. Also, a safety system (with electronic command control) will monitor the parameters of the power supply: if an over-voltage or over-current or ground faults or excessive temperature happens, the chargers will be turned off and the total amount of energy of capacitors will be shorted to the ground by a relay.

During future R\&D programs, the impact of the ground safety on the lifetime of the discharging switches when any faults happen should be studied. Since high +12~kV voltage is applied on the strip-lines and horns, the efficiency of their ceramic electrical isolation under helium atmosphere, high radiation and electromagnetic field has to be studied too. 

\subsection{Thermal dissipation}

The thermal dissipation comes from the heat due to joule effect of the local currents on resistive materials (components, strip-lines, horns) and the energy deposition of secondary particles, especially in the target-horn station. The cooling methods proposed and the power dissipated are given in table \ref{tab:station8}; the dissipations along the transmission lines have been studied by taking into account the skin effect and frequency approach of the current waveform \cite{designreport6}. Water cooling methods will be applied for the 360~kW dissipated from various PSU components in the PSU room and the 208~kW from the horns and the strip-lines placed outside the vessel in the contaminated area. Helium cooling will be applied for the 380~kW dissipated from the targets and the strip-lines placed inside the target-horn station, and finally a small air-climatisation circuit will be applied for the 77~kW from the striplines and other-components in the PSU room. 

\section{Conclusion}
This study describes a feasible and reliable solution in order to provide the 350~kA peak current, 100~$\mu$s pulse at 50~Hz frequency operation needed for the Super Beam horns. That novel solution is based on a modular approach with 8 modules of 4 $\times$ 44~kA for the four-horn system. Each module incorporates a hybrid charger that delivers a high 70~kW rms power at 50~Hz with a precision of $\pm$0.5\%. Each charger is based on the association of a low frequency resonant charger with a high frequency switching one with low electromagnetic interferences. The whole system has a large recovery energy of 97\% obtained by the low values of resistivity. That has been possible by using the method of direct coupled devices with 8 large strip-lines for each horn and with high integration of the module components. The total power for the charging capacitors has also been limited to 560~kW~rms and thus it permits to limit the high operating voltage to +12~kV. The optimal choice of 8 modules of 44~kA connected in parallel have also reduced the critical electrical constraints on all components. A margin of at least 50\% is guaranteed for maximal electrical capabilities of the critical PSU components and allows to attain the goal of 10 years of operation.

\ack We acknowledge the financial support of the European Community under the European Commission Framework Programme 7 Design Study: EUROnu, Project Number 212372. The EC is not liable for any use that may be made of the information contained herein. 

%
%

\end{document}